# A 17 GHz Molecular Rectifier

J. Trasobares[a], D. Vuillaume[a], D. Théron[a], & N. Clément[a,b*]

[a] Institute of Electronics, Microelectronics and Nanotechnology, CNRS, Avenue Poincaré, CS60069, 59652, Villeneuve d'Ascq, France

[b] NTT Basic Research Laboratories, 3-1, Morinosato Wakamiya, Atsugi-shi, Kanagawa, 243-0198, Japan

*nicolas.clement@lab.ntt.co.jp

Molecular electronics originally proposed that small molecules sandwiched between electrodes would accomplish electronic functions and enable to reach ultimate scaling. However, so far, functional molecular devices have been only demonstrated at low frequency. Here, we demonstrate molecular diodes operating up to 17.8 GHz. DC (direct current) and RF (radio frequency) properties were simultaneously measured on a large array of molecular junctions composed of gold nanocrystal electrodes, ferrocenyl undecanethiol molecules, and the tip of an interferometric scanning microwave microscope. The present nanometer-scale molecular diodes offer a current density increase by several orders of magnitude compared to that of micrometer-scale molecular diodes, allowing RF operation. The measured $S_{11}$ parameters show a diode rectification ratio of 12 dB which is linked to the rectification behavior of the DC conductance. From the RF measurements, we extrapolate a cut-off frequency of 520 GHz. A comparison with the silicon RF-Schottky diodes architecture suggests that the RF-molecular diodes are extremely attractive for scaling and high-frequency operation.

**Introduction**

In the field of electronics, there is still a tremandous need for increasing devices speed –more precisely the cut-off frequency $f_T$ keeping the same DC power consumption–, in particular toward the THz frequency range. This frequency domain called the THz gap and corresponding to frequencies between the microwaves and far-infrared, is currently attracting a lot of attention. Electronic devices such as Schottky diodes, resonant tunneling diodes or THz single-photon detectors continue to move into the THz range[1,2].

In that context, molecular electronics, i.e. small molecules connected between electrodes,[3] could play an important role with theoretically predicted transit times down to the fs[4]. Also, the field of molecular electronics is still very active from a basic research perspective[5-26], allowing a rapid progress in devices performance and reliability. The experimental demonstration of a molecular diode as an example of a functional molecular electronic device has generated substantial interest since Aviram and Ratner proposed the pioneering idea of a molecular rectifier[7]. Experimental realizations have been made in both self-assembled monolayers (SAMs)[8-20] or single-molecule[21-25]. Recently, significative progress was achieved based on a better understanding of the intermolecules or/and molecule/electrodes coupling[13-25]. However, so far, molecular diodes have only been demonstrated in the low-frequency regime, a serious limitation from a device perspective. A molecular half-wave rectifier was demonstrated, but the selected frequency of operation was 50 Hz.[14] Diodes based on high mobility organic semiconductors[27] or hybrid materials composed of polymers/inorganic nanoparticles[28] have recently enabled diodes operation at about 1 GHz by optimizing the mobility or charge injection at electrode contacts, but these systems are very different from that of molecular electronics, e.g. electronic transport through small (typically 2-3 nm) molecules between two electrodes. To reach high frequency molecular components, several issues have to be addressed.

First, as pointed out in ref.24, one main issue in these devices is that the conductance has remained extremely low[8-15], seriously impacting the cut-off frequency of the device which could be approximated, in a simple dipole configuration, by $G/(2\pi C)$, with $G$ and $C$ the device dynamic conductance and capacitance, respectively. This limit was recently improved for single-molecule devices by chemical engineering of the molecule/electrode contact using a gold-carbon covalent bond (instead of the usual gold-sulfur-carbon link)[24] but only DC measurements were shown. Also, it is well accepted that conductances per surface area (or per molecule) are orders of magnitude larger in nanometric or single-molecule junctions than in micrometric junctions[29-32] This feature often attributed to surface roughness and the presence of defects. An optimum number of molecules in parrallel per junction has to be found, a single molecule being anyway limited by the quantum of conductance $G_0$=77.5 $\mu$S. A second issue, barely adressed in molecular electronics, is the capacitive circuitry[33-36] for frequency bandwidth optimization. Whereas the low dielectric constant (~2-3)[37] of organic monolayers should be an advantage, fringe capacitances may be dominant when reducing the devices dimension to single or few molecules. Other issues are instrumental: molecular electronics requires statistical studies for convincing demonstrations and it is not simple to implement hundreds of test devices with a top electrode compatible with high-frequency measurements. Also, high frequency measurement systems are mainly designed to measure devices near the 50 Ohm regime, while molecular electronic devices have generally a "high-impedance". Lots of work have been done on RF instrumentation at the nanoscale to overcome this limitation from both academic and industrial side[38-45]. More specifically to molecular electronics, a RF break-junction setup compatible with single-molecule device has been reported for an atomic contact device, but without presence of molecules[46]. AC-STM[47-50], that combines a scanning tunneling microscope (STM) with an AC (alternative current) RF field superimposed on the tip, has been succesfully used to measure fundamental properties of molecules such as the polarizabilities of alkyl chains and $\pi$-conjugated oligomers[48], the resonant oscillations (at 20K) of a molecular chain physisorbed on a surface[49], or the spin resonance of single terbium bis-phthalocyane molecule (at 5K) [50], but no functional device such as a molecular diode has been demonstrated at high frequency and at room temperature.

Here, we demonstrate RF molecular diodes operated up to 17 GHz with an estimated cut-off frequency of 520 GHz. The combination of a large array of single-crystal gold nanoelectrodes (few tens of nanometer in size corresponding to about 150 molecules in the molecular junction) and an interferometric scanning microwave microscope (iSMM[43]) were the key features to perform a statistical study on hundreds of molecular diodes with high dynamic conductances, $G=\delta I/\delta V$ up to 0.36 mS and 110 aF range fringe capacitances. The origin of the large measured conductance is discussed as well as the limiting speed parameters and the perspectives for getting new insights into molecular electronics transport from RF measurements. Finally, based on experimental datas and simple models, we compare the RF-silicon Schottky and the RF-molecular diode architectures. We evidence the interesting perspectives of the molecular electronics approach for scaling and high-frequency operation.

**Results**

**Molecular diode characterization**

Fig.1a shows the studied molecular diode junction composed of a gold nanoelectrode[51-54], an archetype molecule for molecular diodes[13-16] (ferrocenyl undecanethiol : $FcC_{11}SH$), and a Pt tip. The tip curvature radius is large compared to the gold nanocrystal that defines the area of the molecular device (Supplementary Figure 1), leading to a nearly ideal parallel-plate structure with few nm diameter[51,52,54]. In this study, the probe (Pt tip) is connected through a Bias-T to an interferometer for RF reflectometry measurements (iSMM[43]), and to an amperemeter (Resiscope) to get simultaneously the DC current (Supplementary Figure 2). Chemical analysis are performed on a large (1cm x 1 cm) array with billions of the gold nanocrystals thiolated with $FcC_{11}SH$ (Fig.1b), and they confirm thgrafted grafting of molecules on the nanocrystals. First, the X-ray photoelectron spectroscopy (XPS) spectrum (Fig.1d) illustrates an excellent agreement with earlier studies. It shows a Fe doublet located at 707.8 eV () and 720.7 eV[55], and a ferricenium peak at 710.6 eV[56] and 723.9 eV. While not always observed, e.g. in large area self-assembled Fc monolayers on metal surfaces[55], the presence of the ferricenium peak (Supplementary Figure 3 for details) could be related to the silicon dioxide between dots in the presence of moisture. It may affect the stable charge state of Ferrocene molecules at gold nanocrystal sides (Fig.2d, inset) , dots borders representing ~80% of the surface area (Supplementary Figure 4). Second, electrochemical characterization by cyclic voltammetry (Fig.1e) shows one broadened peak—more specifically two peaks— appearing at E= 0.34 and 0.37 V (vs Ag/AgCl), in the range of expectation for Fc molecules[13]. Peak broadening or the presence of two peaks is usually related to molecular interactions in a densely packed SAM [13,15,57] The average number of molecules per dots (~860 considering the total dot surface –top plateau and sides– for these 20nm-diameter, 3nm-thick truncated cones[51]) corresponds to an average area per molecule of 0.405±0.04 nm$^2$. It is assessed from the total charge measured by cyclic voltammograms at different sweep rates, and with consideration of an uncertainty on the averaged nanocrystal area (Supplementary Figures 4,5 and supplementary Table 1). This averaged value is 10% larger (e.g. molecule density slightly smaller) than for an ideal, fully packed, SAM with 0.37 nm$^2$ per molecule[57-59], but in the range of expectations for a dense SAM (Supplementary Note 1). Considering the sole top surface of the truncated cone (Supplementary Figure 4), we can estimate to ~150 the number of molecules contacted between the iSMM tip and the gold nanocrystal (molecules on dot sides do not contribute to the electron transport). Fig 1f shows a DC current-voltage, *I-V,* histogram obtained over 100 junctions from scans at different tip bias (bias applied on the tip) – see details in SI.

The DC rectification ratio ($R_{DC}=|I(V=1V)/I(V=-1V)|$ ~ 5) is in the range of usually reported values for many other types of molecule diodes [11-24], albeit not the largest one reported for the same molecule[15]. The maximum dynamic conductance, $G=\delta I/\delta V$, calculated from the data in Fig.1, is 360 $\mu$S at 1V. It corresponds to a normalized conductance per molecule of 2.4 $\mu$S if we suppose that all 150 molecules are contacted, and is we neglect molecule-molecule interactions which are known to increase the total conductance of the junction compared to the sum of its parts[60,61]. Although large, it remains more than one order of magnitude below that of the quantum conductance (77.5 $\mu$S). With this in mind, we propose a simple energy band diagram[9] of the molecular device in Fig.2. The energy level for the HOMO is estimated from cyclic voltammetry to be -5.05 $\pm$ 0.05 eV relative to vacuum (Supplementary Figure 6 and Supplementary Note 2). In this situation, a higher conductance is expected for positive bias on the Pt electrode when the HOMO levels of the Fc group are pulled down to lie in the energy window defined by the Fermi energy levels of the electrodes (resonant tunneling). Note that the direction of rectification is opposite to that of other published results with same/similar molecule.[16-20] This feature can be mainly attributed to the large metal work function of the Pt top electrode (Supplementary Table 2 and Supplementary Note 3) and to the work function difference between the two electrodes (Au and Pt), which favor a rectification direction in the opposite direction of the one attributed to the energetics of the Fc molecule in the junction.

The dynamic conductance (up to 0.36 mS at 1V) is suitable for high frequency operation as discussed later in the paper. The capacitance will be discussed together with RF measurements.

**RF-molecular diodes**

Figure 3, panels a, b and c, show the iSMM images of DC current, RF reflection coefficient $S_{11}$ amplitude ($|S_{11}|$) and phase ($\varphi S_{11}$) collected simultaneously at 0.8 V and 3.78 GHz. $S_{11}$ is the ratio of reflected RF signal power over the incident RF signal power. $|S_{11}|$-$V$ and $\varphi S_{11}$-$V$ 2D histograms are

shown in Figs.3d,e. A molecular diode behavior is clearly observed with a rectification ratio in $|S_{11}|$ at +/- 1 V, $R_{RF}=|S_{11}|_{V=1V} - |S_{11}|_{V=-1V}$ of 12 dB, whereas for a symmetric molecular junction (alkyl chains of similar length), no rectification was observed (Supplementary Figures 7,8). The molecular diode device is still operating at a frequency of 17.8 GHz, the upper limit of our experimental setup (see Figs.3f-h for the *I-V*, $|S_{11}|$-*V* and $\varphi(S_{11})$-*V* at 17.8 GHz), although we notice a reduced rectification ratio in $|S_{11}|$ at +/-1V of 4dB. We do not notice any increase in the DC current induced by the RF signal (photocurrent). Fig.3f shows the measured DC *I-V* with/without RF applied on the tip (black solid line), confirming that no photocurrent is measured, even at 17.8 GHz. This result is not surprising at room temperature as the RF excitation corresponds to an energy of 73.6 µeV, much smaller than the thermal energy (25 meV). Also, the estimated absorbed power (<1 µW) is typically lower than Joules power (up to few µW). Eventually, we could have expected a RF-modulation of interfacial dipoles[48], observable either on the current (through the so-called molecular gating[54,62]) or on $S_{11}$ via a capacitance change due to a polarization effect[35,48,63]. These effects are not observed due to the weak dipoles and the presence of a fringe capacitance, which hides the effects. The rectifying behavior on $S_{11}$ can be related to the DC properties. Considering the RC equivalent circuit shown in Fig.4a (the dynamic conductance *G* must be derived from $\delta I/\delta V$ of the DC current-voltage curve or from the RF measurements as detailed below, the capacitance is $C=C_{mol}+C_p$ with $C_{mol}$ the capacitance of the molecular junction and $C_p$ the fringe capacitance between the tip and substrate), the RF reflection coefficient $S_{11}$ of the device under test is related to the impedance of the measured device, $Y=G+jC\omega$ by[43,45]:

$$S_{11} \simeq -2Z_c \cdot A_0 \cdot (Y-Y_0) \quad (1)$$

where the parameter $A_0$ takes into account losses, gain and shifts due to cables, passive and active elements and $Y_0$ is the admittance for which fully destructive interference occurs ($S_{11}=0$), and $Z_c$ is the characteristic impedance of 50$\Omega$. To extract the RF conductance, $G_{RF}$, from the $S_{11}$ measurements (Fig.3), we need a calibration protocol (Supplementary Methods 1 and Supplementary Figure 9) to determine the parameter $A_0$, $Y_0=G_0+jC_0\omega$, and the capacitance $C$. From this calibration protocol, we get $A_0$=(61dB;-153°) at 3.78 GHz and (52dB; -90°) at 17.8 GHz, $G_0\approx0$, $C_0$=300 aF, and $C$=110 aF. Using these values in Eq.1, $G_{RF}$-$V$ curves (Fig.4b and c) are extracted from $S_{11}$ data (Fig.3) at 3.78 and 17.8 GHz, and compared with $G_{DC}$-$V$ curves. We observe a reasonable agreement both at 3.78 GHz and 17.8 GHz. This means that this molecular diode works without significant perturbation up to 17 GHz. Although our experimental setup system is limited to 18 GHz, we can estimate the cut-off frequency $G_{max}/(2.\pi.C)$=520 GHz, where $G_{max}$ is the maximum dynamic conductance (0.36 mS in Fig.4b at 1V) and $C=C_{mol}+C_p$~110 aF as discussed above. Note that $C$~$C_p$ as $C_{mol}$ is expected to be of only few aF from geometric dimensions and a dielectric constant of 2-3 for the organic monolayer[37].

## Discussion

The present nanometer-scale molecular diodes offer a current density increased by more than nine orders of magnitude compared to that of micrometer-scale molecular diodes with the same molecule[15,16], allowing RF operation (see Supplementary Table 2 for a direct comparison of current densities for various architectures[16-20]). As mentioned in the introduction, this difference can be mainly explained by the fraction of molecules connected due to a lower roughness in nanometric junctions[Error! Bookmark not defined.-33]. In our system, gold nanocrystals have an atomically flat top surface[54]. Also it is possible that the π-orbitals of the ferrocene molecule overlap with the Pt spill-over electron density, resulting in a large contact conductance, even without chemical contact[6,9,64,65]. If considering an upper conductance limit of $G_0$ for the dynamic conductance per molecule, then at least 5% of the molecules contribute to the electronic transport (e.g., probably a large fraction of the molecules are connected). Nonetheless, we also believe that the molecular packing plays a role as the alkyl chains, that govern the coupling strength between the Fc molecules and the bottom electrode[9], take a small space compared to that of Fc heads[15,16,57]. Cyclic voltammetry experiments (Fig.1e and Supplementary Note 1) suggested that the present monolayers on the gold nanocrystals are 10% less dense than for a fully packed monolayer. Weaker van der Waals forces between alkyl chains should reduce the distance between ferrocene molecules and gold nanocrystals[66] (e.g. increase the tunneling rate), reduce the tunneling decay ratio[16], and weaken the effect of tip load[54] which we observed experimentally (Supplementary Figures 10-12 and Supplementary Note 4 for a discussion on the effect of tip load[67] and molecule). Finally, we cannot exclude an impact from the metal atom oxidation state on the conductance[68] (e.g. here the presence of Ferricenium molecules seen on XPS spectra). However, if located on nanocrystals sides (Fig.1d, inset), ferricenium molecules should barely affect the electronic properties. Moreover, T.Lee and coworkers[16] suggested that electron transport through ferricenium is slightly lower (a factor ≈ 1.5) than through ferrocene, thus the possible presence of few ferricenium ions on the top of the nanodot is not drastic.

In the present experiments, the change of frequency from 3.78 GHz to 17.8 GHz already points out the appearance of the capacitive contribution $|C-C_0|/\omega$ as a plateau in $|S_{11}|$ at low bias. Theoretical predictions suggest that an additional inductive contribution[4] (inertia related to carriers effective mass) or oscillation in $S_{11}$ due ferrocene spinning[69] should be observed in the THz range. Recently, Tan et al.[70] have shown that tunneling charge transfer plasmon modes can operate through molecules at hundreds of THz, suggesting that there might not be physical limits to operate molecular rectifiers in the THz gap. However, in the hypothesis of incoherent tunneling, the limiting speed for capture/emission of a charge by a redox molecule[13] or more generally from a trapping site under large electric field[71] still remains to be explored. The present studied frequency range should already be ideal for routine measurement of the shot noise from $S_{11}$ parameter[72], another degree of freedom for getting insights in the operation mechanisms of molecular diodes or other molecular devices as it is directly related to the transmission function in the Landauer formalism.

We now compare the RF-silicon Schottky and RF-molecular diodes architectures (Figs.4d,e). Schottky diodes are a key component in RF mixers, and are used in wifi or mobiles phones. The cut-off frequency $f_T$ is typically the figure of merit, not only for the maximum frequency operation, but also for optimized power consumption in practical applications[73,74]. To optimize the conductance, RF-Schottky diodes are biased in the non-linear regime (induced by a depletion layer at the metal/semiconductor interface) just below the linear regime governed by the spreading resistance in the substrate ($R_S$). A thin undoped epitaxial layer (few nms) enables to keep the highly doped substrate as close as possible to the interface while not getting a too large junction capacitance $C_J$ (a typical value[73] for $C_J/A$ is 6.2 $\mu$F/cm$^2$). $f_T$ is usually considered as $1/(2\pi R_S.C_J)$[73,74]. Whereas $C_J$ scales as the junction area $A$, $R_S$ scales as $A^{-1/2}$ because the substrate thickness is large compared to that of the junction diameter $d$[74]. This has led to smaller and smaller RF-Schottky diodes with $f_T$ scaling as $A^{-1/2}$ (green line Fig. 4f). A $f_T$ of few THz was demonstrated for a 250 nm diameter diode[73], close to the theoretical prediction (Fig.4f). However, further scaling these diodes should induce unsustainable large current densities (the typical upper range of failure current limitation is ~3x10$^8$ A.cm$^{-2}$)[75,76], unless a small oxide layer at the contact interface increases the resistance[77]. A maximum current density imposes $R_S$ to scale with $A$, leading to a saturation of $f_T$ at $d$ below 50 nm (Fig.4f).

In the RF-molecular diode architecture (Fig.4f), organic molecules replace the undoped semiconductor epitaxial layer. The theoretical cut-off frequency limit is given by $1/(2\pi R_S.C_{mol})$, resulting in a gain of $C_J/C_{mol}$~5.6-8.8 when compared to Schottky diodes (typical $C_J$ is considered[73]) thanks to a small dielectric constant of 2-3 for organic monolayers[37] (blue curve in Fig.4f). The failure issue at high current density remains valid in the molecular rectifier architecture (see SI, section 7). However, in the present experiment, as $G$ is lower than $1/R_S$ in the +/-1 V biasing range (current governed by molecules conductance), such limit is not reached with these 20nm-diameter gold nanocrystal electrodes (although not so far). The comparison of the experimental values at 1V for $R_{mol}=1/G$ and $f_T$ with theoretical optima suggests that the actual limiting factor is not the molecule conductance, but rather fringe capacitances that degrade $f_T$ by about two orders of magnitude. Still, improvement of the device conductance (optimization of the conductance per molecule[24] and rectification ratio (e.g. a larger rectification ratio has been reported with the same molecule using other electrodes[15]) by tuning the metal work function or molecular organization could enable reducing the optimum operation voltage while keeping the same $f_T$.

Finally, we stress that for practical applications, nano silicon Schottky diodes are assembled into arrays so as to benefit from the high performances related to small dimensions while keeping $R_s$ to few Ohms[73] (close to the 50 $\Omega$ impedance). By analogy, standalone optimal RF-molecular diodes could be obtained in the future based on the present array of molecular nanodiodes covered by a top contact. Such top contact electrode could be, for example, graphene ribbons thanks to their RF compatibility[78], small contact resistance[79], and rigidity so as to decrease fringe capacitances between dots.

To conclude, we demonstrate a high-frequency molecular rectifier composed of self-assembled small molecules. It brings new hopes and perspectives for a chemistry-based electronics. There is a tradeoff between the conductance, the function (here rectification) and capacitances. The present configuration was optimized to demonstrate a 17 GHz molecular rectifier with a 520 GHz cut-off frequency: The ~150 molecules sandwiched between a gold nanocrystal and a Pt tip electrode enables to get a ~0.36mS dynamic conductance while keeping a small capacitance in the 100 aF range. The sole reduction of the fringe capacitance to 50 aF by using an ultra-sharp tip[45] or by surface functionalization between nanocrystals would enable to operate these diodes in the THz range where new functionalities are theoretically predicted, e.g. a huge increase of the dynamic conductance of the molecular devices.[80,81]

**Methods**

**Gold nanodot electrode fabrication**

The fabrication of the gold nanocrystal arrays have been described elsewhere[51]. For e-beam lithography, we used an EBPG 5000 Plus (Vistec Lithography©). Silicon (100) substrates of 0.001Ω.cm resistivityThe were exposed to UV-ozone and the native oxide removed by HF at 1% before resist (PMMA 950K, diluted 3:5 with anisole, 45 nm thick)spin-coating..

The e-beam acceleration voltage was 100 keV. The dose per dot corresponds to 3-4 fC[82]. Just prior evaporation, the native oxide was removed with dilute HF solution to allow good electrical contact with the substrate Similarly, we stress that it is important not to use any adhesion layer such as Ti as the diffusion of gold in the silicon substrate is required to get the Au nanocrystal structure[51]. used.

After thermal annealing at 260°C during 2 h under $N_2$ atmosphere, we get Single crystal Au nanodots. (Supplementary Figure 1). During this process a thin layer of $SiO_2$ covers the dots[51]. It is etched by HF (1%, 1mn) before SAM grafting.

**Self-assembled monolayers**

For the FcC$_{11}$SH monolayer, we used 11-ferrocenyl-1-undecanethiol of 95% purity from Aldrich. The gold nanocrystals were exposed to a solution of 20% dichloromethane and 80% ethanol (VLSI grade from Carlo Erba)during typically 24 h in a glovebox in the darkness. The treated substrates were rinsed in ethanol and further cleaned with gentle ultrasonication (20% power, 80 kHz) in chloroform (99% from Carlo Erba) during 1 min. The gentle ultrasonication was required to avoid pollution of the tip from adsorbed molecules on silica. In the future, ultrasonication may not be necessary if a monolayer of silane molecules is grafted between dots to prevent non-specific adsorption. Similarly, recent studies performed by Jian et al.[83] suggest that a fully packed SAM could be obtained after an additional purification step to avoid for example the presence of disulfides. We have not done such extra purification step in the present study. It could be an interesting perspective to evaluate its impact on the performance of RF molecular diodes.

**XPS measurements**

XPS measurements have been performed using a monochromatic Al (Kα) X-ray source (1486.6 eV) and an analyzer pass energy of 12 eV. The spectrometer was from Physical Electronics (5600). A

The analyzer acceptance angle, the detection angle, and the analyzed area were set to 14 degrees, 45°and 400 µm, respectively.

**iSMM and resiscope set-up.**

The interferometric scanning microwave microscope (iSMM), recently developed at IEMN[43], is an adjustable interferometer analogous to a Mach-Zehnder configuration. It consists on a coaxial power divider and two coaxial hybrid couplers associated to an active variable attenuator (Supplementary Figure 2). The VNA source delivers the incident signal, $a_1$, that is split into two parts. One part of the signal, used as a reference, is adjusted in magnitude by the variable attenuator in order to adjust the interference; the second part feeds the AFM tip through a coupler and is then reflected back by the device under test (DUT). Both signals are combined inside the second coupler and then the signal, $a_3$, is amplified and analyzed the VNA receiver.

This iSMM is connected with an amperometer by a bias-T that allows electrical characterization of the DC current simultaneously (Supplementary Figure 2). As an independent measurement, the ResiScope (log amplifier) measures the sample resistance through the High Performance Amplifier (HPA).

The iSMM measurements were done with Agilent 5600 AFM combined with an Agilent PNA series network analyzer. The cantilevers from Rocky Mountain Nanotechnology area Pt solid wires glued to an alumina chip to enable compatibility with high-frequency measurements. The tip used (RMN 12Pt300A) were 300 µm-long, 60 µm-wide and a shank length of 80 µm with a tip curvature radius of about 20 nm. In practice, we found that the curvature radius for these Pt wire tips was about 80 nm[45], so we had to increase the interdot distance to 200 nm. The spring constant is given to be about 0.8 N/m enabled to control force and set it to 18 nN (see Supplementary Note 4 for a discussion on the tip load). Because the dots dimension were larger in that study when compared to refs.52,54, the influence of the force was reduced (smaller force per surface unit) in the nN range. Images were acquired with a sweep frequency of 0.1 lines/s (go and back) and 512 pixels/line. For an image of 15 µm$^2$ it corresponds to a scan rate of 3.12 µm/s. Given an interspacing between nano-junctions of 250 nm, we can measure 3600 junctions in 1h 22 min at a given DC bias. The voltage applied on the substrate by the Resiscope software, but for convenience and comparison with other published data, all the voltage values are reported in the paper figures as applied on the tip.

**Cyclic Voltammetry**

Fig.S4-a shows the electrochemical cell used in this study. The 0.5 mL container is filled with $NaClO_4$ (0.1M in water) as the electrolyte (Supplementary Figure 13). The cell is connected with the Solartron ModuLab potentiostat by the three typical electrodes. Typical Ag/AgCl and Pt wire electrodes were used as the reference and counter electrodes, respectively. The working electrode is the gold nanocrystal array on the silicon substrate.

Before the experimental measurements, the electrochemical cell was cleaned with ethanol and DI water. "Test" sweeps between -0.1 and 0.6 V with a highly doped silicon substrate (without dots) are measured to confirm that there is no peak due to contamination. Cyclic voltammetry (stable under several voltage cycles) proves the presence of ferrocene-thiol electroactive molecules and allows their quantification.

**Image treatment**

Images were treated with Origin C program with 2 functions. The 1st function applies a threshold to remove the background noise. Then, 2nd function obtains the maximum per dot by checking the nearest neighbors. Due to temperature fluctuation in the interferometer, or to the mechanical relaxation in the cables, both active and passive elements on the interferometer may undergo a particular drift. The interference set at the beginning of the experiment may moves as we see in the Supplementary Figure 14a and it causes changes on our measures as noticed on Supplementary Figure 14b (change of the background color). There are two contributions to this drift: One is related to the increment of the reflected signal when the interference moves from the selected frequency. It is solved by subtracting the ground level measured on the native $SiO_2$. The second contribution comes from the fact that the reflected signal decay exponentially when the interference moves away the selected frequency. The procedure followed to overcome this limitation during the measurements of the $S_{11}$ parameters vs. voltage has been to reduce the image size to about 100 nanojunctions and normalize the data to those obtained on the flat zone. This procedure has been validated for inorganic nanocapacitors using an on chip calkit[45]. Therefore, the complete curves with 41 images at different bias can be done in only 3 hours (Supplementary Figure 15), which significantly reduced drift issues.

**Data availability statement**

All relevant datas are available from the authors on request.

**Acknowledgments**

The authors would like to thank T. Hayashi, A. Fujiwara, and T. Goto from NTT Basic Research Labs for fruitful discussion and support, Xavier Wallart and D. Guerin from IEMN for discussions on XPS, and E. Tournier from IES Institute for discussion on THz devices. J.T. thanks PhD funding by a Marie curie grant, EU-FP7 Nanomicrowave project. We acknowledge support from Renatech (the French national nanofabrication network) and ANR Excelsior Project.

**Authors contributions**

J.T. fabricated the devices, performed the electrochemical and electrical measurements, treated the datas and performed fits to $S_{11}$ parameters. D. T. brought continuous input to the SMM experiment, wrote the fitting program and performed fits to $S_{11}$ parameters. D. V. provided expertise in molecular electronics and molecular diodes and inputs to the manuscript. N.C. supervised the study and wrote the paper. All authors discussed the results.

**Additional informations**

Additional information

Supplementary Information accompanies this paper at http://www.nature.com/naturecommunications

Competing financial interests: The authors declare no competing financial interests.

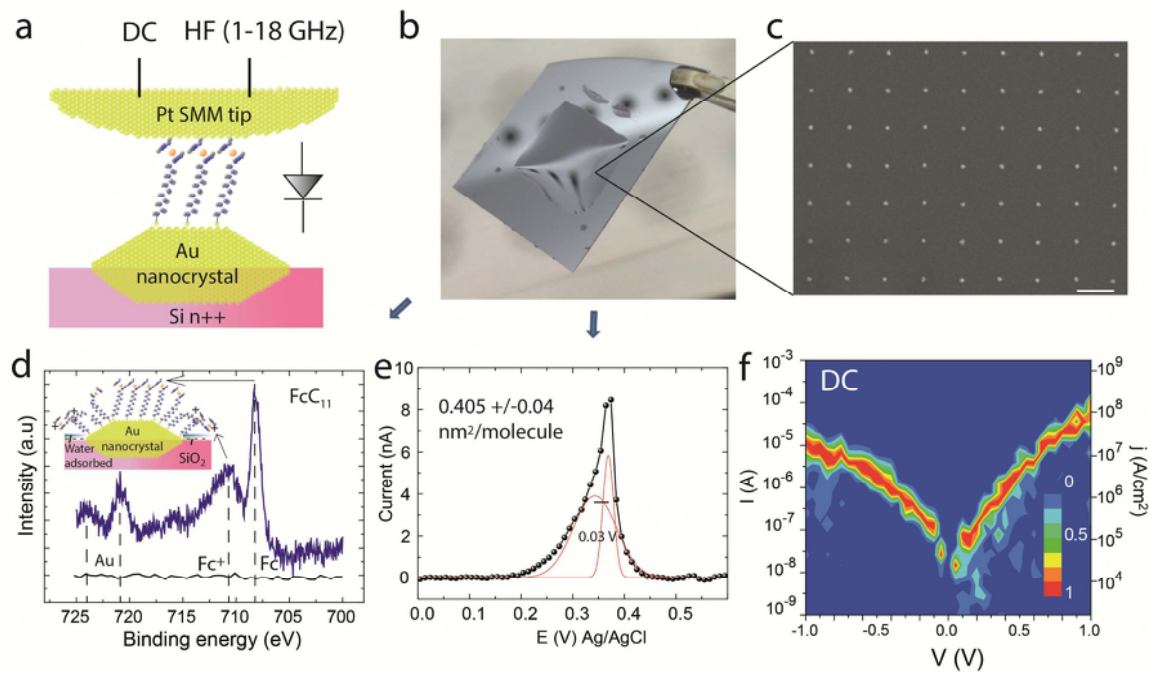

**Figure 1 Description/characterization of the RF Molecular rectifier**

**a** Schematic representation of the molecular junction composed of a gold nanocrystal, Ferrocenyl undecanethiol (FcC$_{11}$) molecules enabling rectification properties, and a Pt tip. Nanocrystals form an ohmic contact to a highly doped silicon substrate. The Pt tip is biased (through a bias-T) to both DC and HF (1-18 GHz) excitation simultaneously. **b** Picture of a 1cmx1cm array of gold nanocrystals used for X-ray photoemission spectroscopy (XPS) or Cyclic Voltammetry measurements. The picture is taken just after dipping the sample into HF (for removal of the SiO$_2$ covering dots), the gold nanoarray area being identified through an hydrophilic/hydrophobic contrast. **c** Gold nanodot array imaged by Scanning Electron Microscope (SEM). The scale bar is 200 nm. **d** XPS measurements for SAMs of FcC$_{11}$ grafted on gold nanocrystals (~1 billion dots fabricated by high-speed lithography) showing the presence of a Fe doublet related with ferrocene at 707.8 eV and 720.7 eV (Fe $2p_{3/2}$ and Fe $2p_{1/2}$, respectively), anda Fe doublet related to ferricenium at at 710.6 eV at 723.9 eV ($2p_{3/2}$ and $2p_{1/2}$, respectively). The XPS signal for the bare Au nanoarray is also shown as a reference. Inset: schematic representation of the SAM with Ferricenium molecules located at dots borders due to the presence of a negatively charged silica. **e** Cyclic voltammetry measurements supports the presence of ferrocenyl molecules on the nanodots with a double peak at *E*= 0.34 V and 0.37 V versus Ag/AgCl as a reference electrode in agreement with previous studies[15]. **f** 2D histogram (normalized to 1) showing the *I-V* (and *J-V*) curve from one hundred junctions on 20 nm gold nanoparticles. The voltage step is 0.1 V, and the 2D histogram is obtained by the contour plot function (Originlab©). The applied load was 18 nN (see Supplementary Note 4 for a detailed discussion on the tip load).

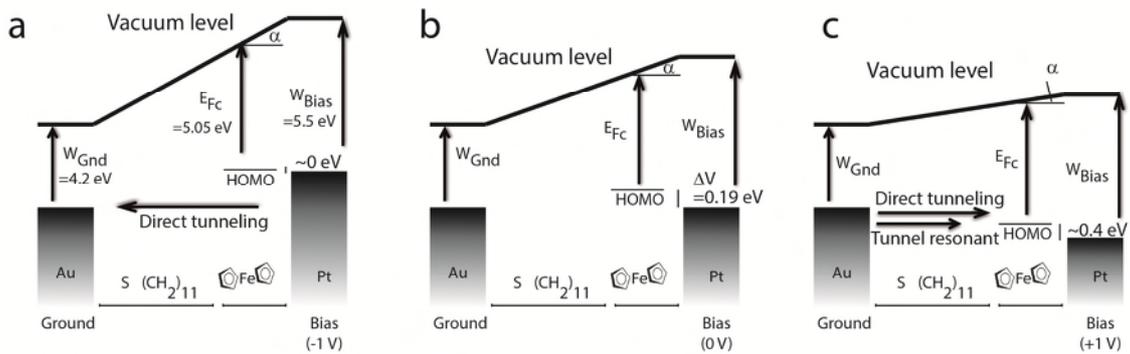

**Figure 2 Sketch of the energetic level of the molecular device**

**a** Sketch of the energetic level of the molecular device when negative (-1 V) is applied on the Pt tip (see details in SI). **b** Same as **a** with 0 V applied on the tip. **c** Same as **a** with +1V applied on the tip. We considered a coupling parameter of 0.8 to the Fc from the gold atom (80% of the potential drop occurs in the alkyl chain) to calculate the energy shift of the HOMO levels in the junction (see Supplementary Figure 6 for details).

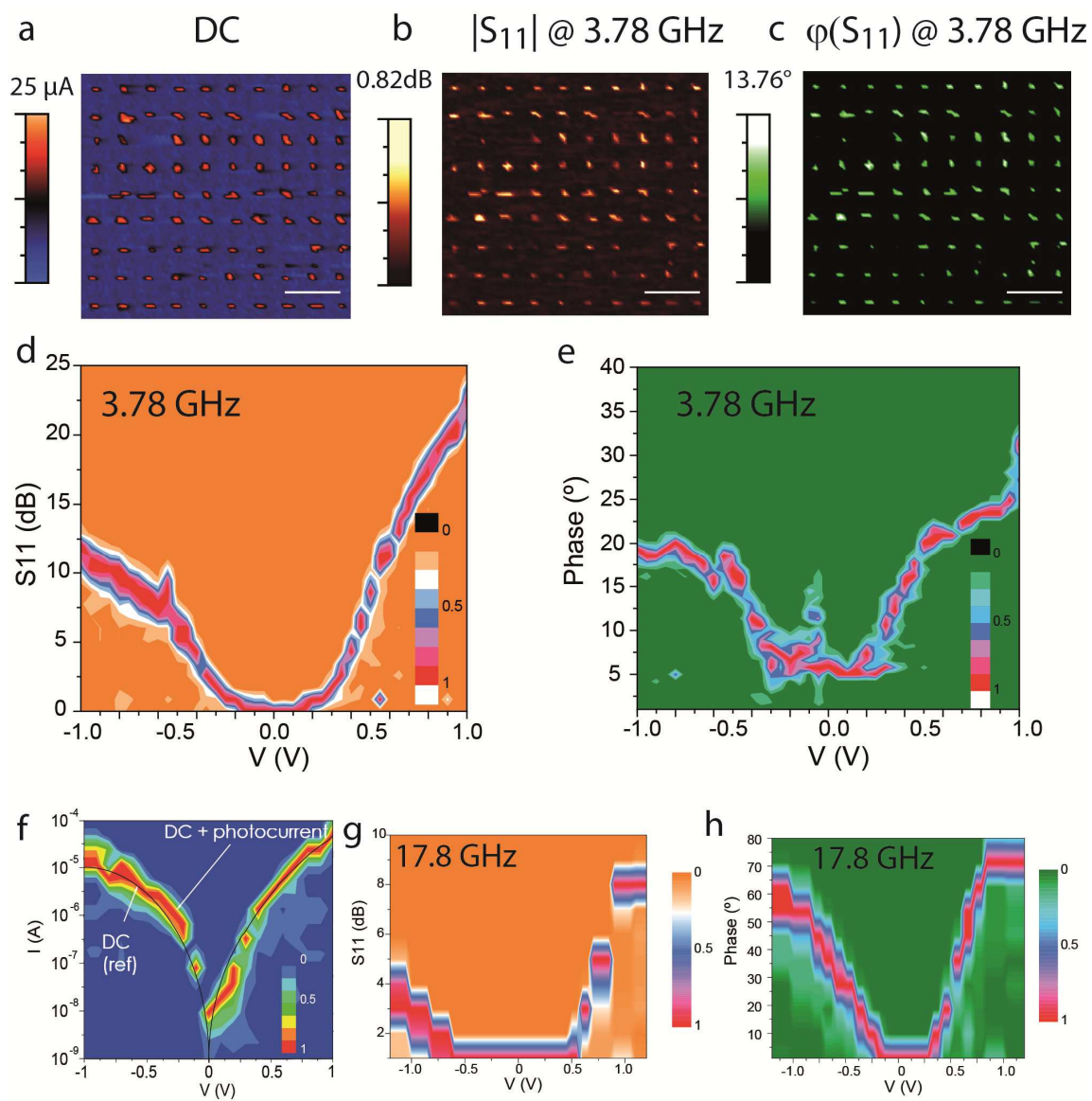

**Figure 3 Demonstration of a molecular rectifier at 3.78 and 17.8 GHz**

**a-c**. iSMM images at 0.8 V DC and 3.78 GHz measured simultaneously by the amperometer (Resiscope) for the DC current (a) and VNA (vectorial network analyzer) for amplitude (b) and phase(c) $S_{11}$ parameters. Scale bar is 500 nm. **d** and **e** 2D $|S_{11}|$ histogram (normalized to 1) vs. tip bias (V) and $\varphi(S_{11})$-V 2D histograms (normalized to 1) generated from one hundred molecular rectifier junctions. The voltage step was 0.05 V and the contour plot generated automatically (Originlab©). The applied load was 18 nN. **f** 2D DC *I-V* histogram from one hundred of ferrocenyl undecanethiol gold nanojunctions with a 17.8

GHz RF input signal. The DC reference current (solid line) when no RF input signal was added is shown for comparison. It was obtained from the average *I-V* from a 2D histogram, without RF power. The voltage step was 0.1 V and the contour plot generated automatically (Originlab©). **g-h** 2D |$S_{11}$| vs voltage curve from one hundred of ferrocenyl undecanethiol gold nanojunctions at 17.8 GHz and related φ($S_{11}$) vs.V curve. The voltage step was 0.1 V and the contour plot generated automatically (Originlab©).

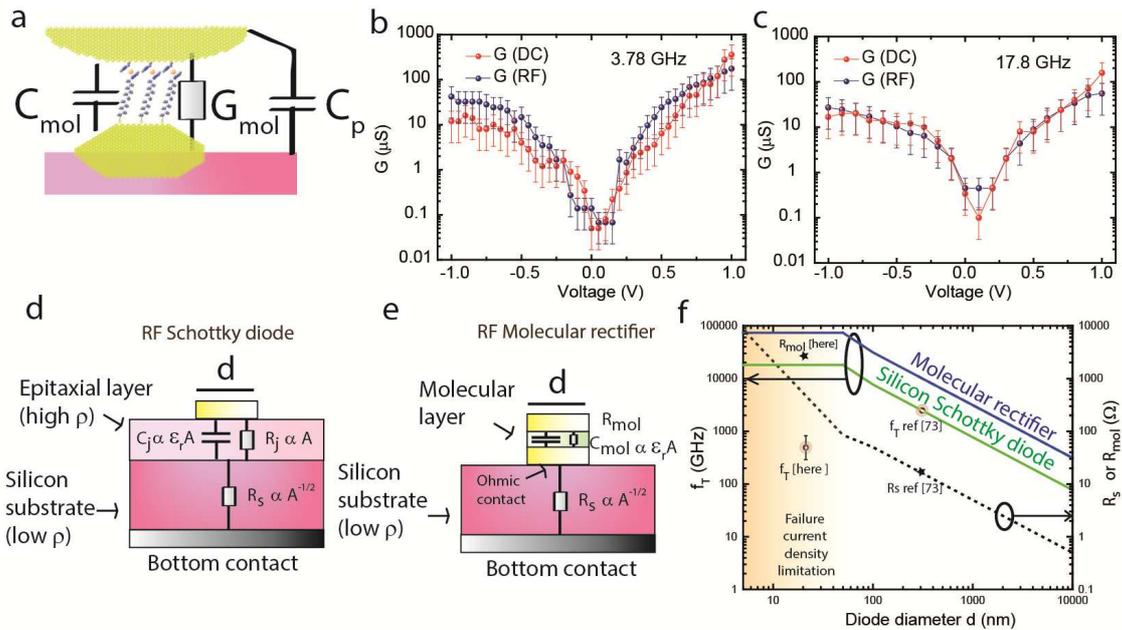

**Figure 4 Perspectives for RF-molecular rectifiers**

**a** Equivalent circuit representation of the device. Molecules can be decomposed as a conductance in parallel with a capacitance ($G_{mol}$ and $C_{mol}$). Also, a fringe capacitance ($C_p$) between the iSMM tip and the sample has to be considered. **b** Conductance $G_{mol}$ estimated from both the DC measurement ($\delta I/\delta V$) —red curve—(technically obtained after multi-exponential fit with 200 points of the DC I-V curve: 21 points), and from $S_{11}$ parameters (Eq.1) —blue curve— (see SI, section 8 for fitting details). The error bar in log scale is considered to be the same as that of full width half maximum in current histograms. **c** Similar curves as in b at 17.8 GHz. The error bar in log scale is considered to be the same as that of full width half maximum in current histograms. **d** Schematic cross section of the RF Schottky diode architecture. The high resistivity epitaxial layer is thin (few nm) so as to to tune $R_j$ up to $R_s$, but not too thin to avoid a large $C_j$. The substrate is highly doped (resistivity $\rho_s = 0.001\Omega.cm$). Its resistance scales as $A^{1/2}$ where A is the junction area. **e** Schematic cross section of the proposed RF molecular rectifier. The molecular layer plays the role of the diode with a small dielectric constant $\epsilon_r$. Similar to Schottky diodes, the molecular diode is connected to a highly doped silicon substrate. **f** Graph illustrating the theoretical (ideal) $f_T$ and resistance ($R_s$ or $R_{mol}$) for both the RF-molecular rectifier and the RF-Schottky diode architectures shown in d,e. The dash curve corresponds to $R_s=\rho_s/2d$ with $\rho_s$=1mΩ.cm[74]. $C_j/A=6.2\mu F/cm^2$ from refs. 73,74 and $C_{mol}/A$=0.9-1.4 $\mu F/cm^2$ based on a dielectric constant of 2-3 for the monolayer and a monolayer thickness of 1.9 nm (the length of the molecule). A current density failure limitation of $3\times10^8$ A/cm$^2$ from refs 75,76, induces a saturation of $f_T$ in the graph. Measured $R_{mol}$ at +1V and estimated $f_T$ are also indicated. The error bar is related to the conductance dispersion from 2D *I-V* histograms.

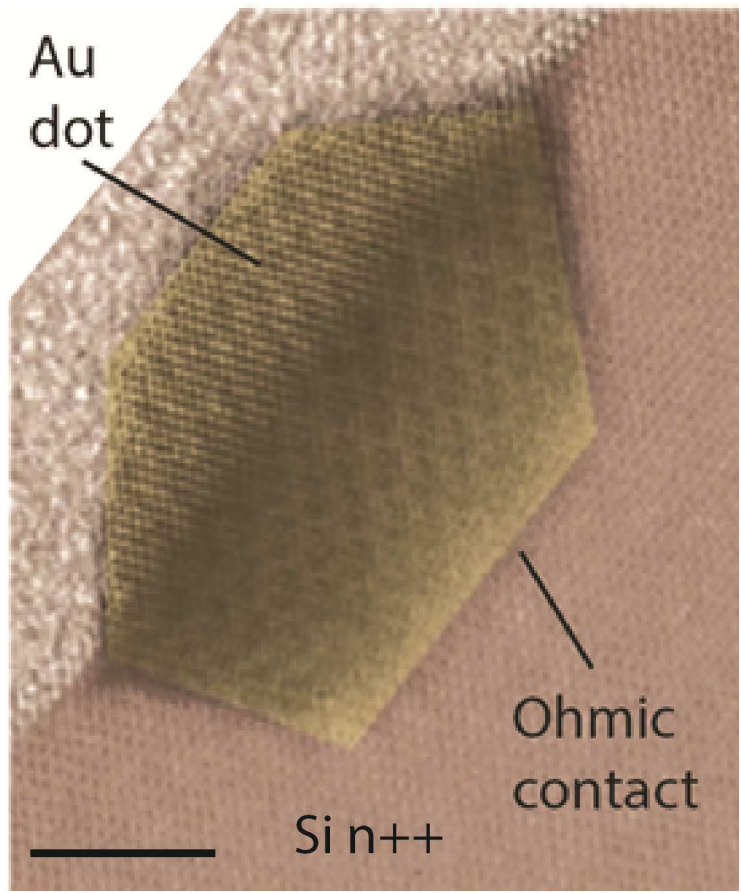

**Supplementary Figure 1.** Gold nanodot electrode fabrication Coloured TEM images of an Au nanodot (adapted from Ref.1). Scale bar is 5 nm.

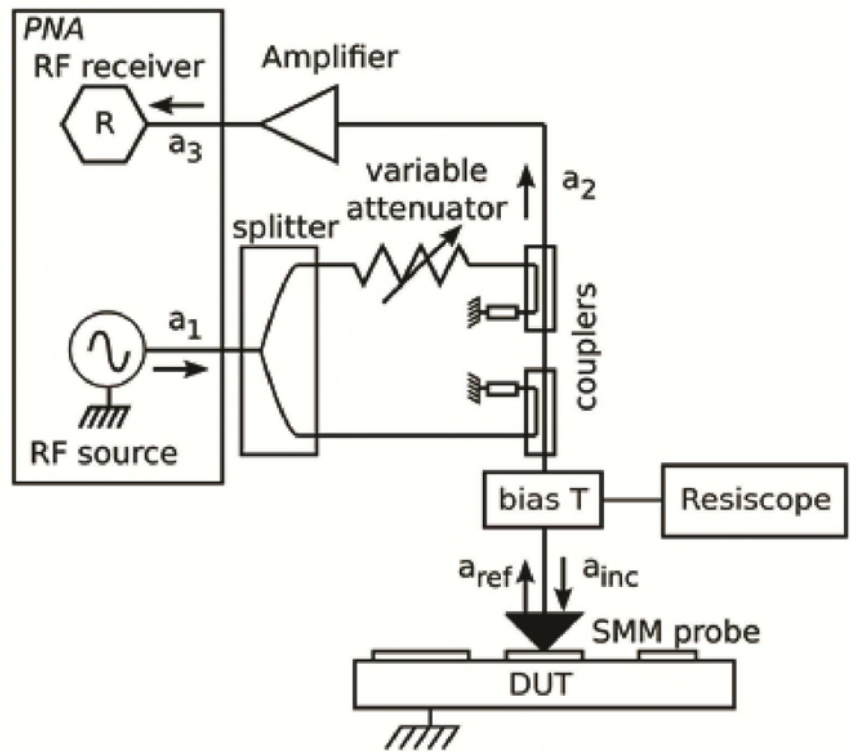

**Supplementary Figure 2**. Experimental setup composed of an interferometer[2] and log current amplifier (resiscope[3])

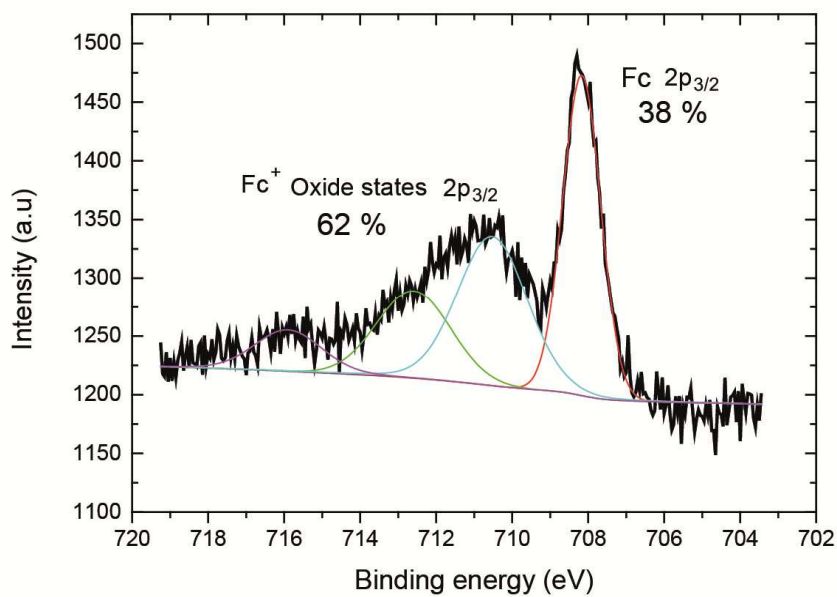

**Supplementary Figure 3.** XPS spectrum (zoom on Fig.1d) used to evaluate the relative fractions of Ferrocene (38%) and Ferricenium=62% from respective areas of each Gaussian fit. The estimation for the area on top (21%) and sides (79%) could match with ferricenium molecules mainly located on dots sides. When compared to ref[4], the FWHM is halved for both peaks, but the ratio of FWHM for both peaks is similar (e.g. broad peak for ferricenium). Background was substracted by Shirley algorithm[5].

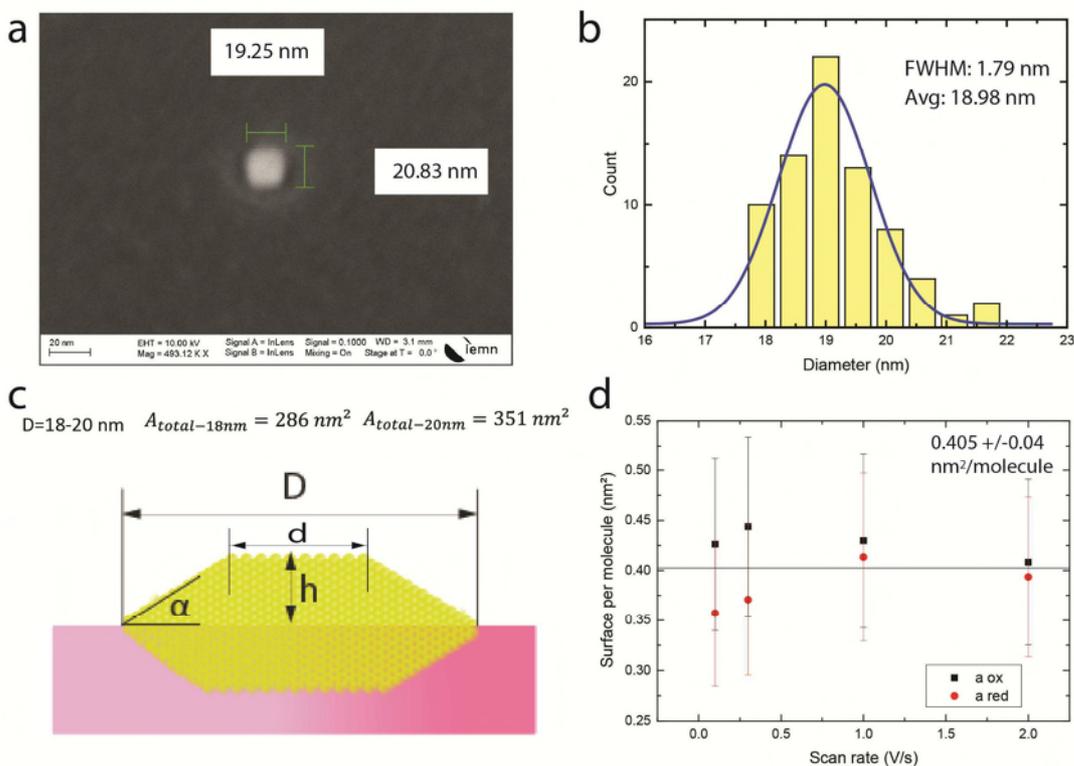

**Supplementary Figure 4 a** Example of SEM image for a gold nanocrystal. **b** Based on the estimation of nanocrystals diameters from 80 images as **a**, we obtained the statistical repartition of 19 nm +/- 1 nm. **c** Schematic cross section representation of a nanocrystal from Fig.S1 used to estimate the total area of a nanocrystal. D is the total diameter, d is the diameter on top of the dot, $\alpha$ was considered to be 30° and h=3 nm. The area was considered with the following formula: $\pi/4*(D-2*h/\tan\alpha)^2+H*(D-h/\tan\alpha)*h/\sin\alpha$, the first term corresponding to the area on top and the second term to the area on sides. **d** Extracted surface per molecule for each scan rate (Supplementary Figure 5). The error bar is related to the uncertainty on dot diameter estimated in **b**. The averaged area per molecule is 0.405 +/- 0.04 nm²/molecule.

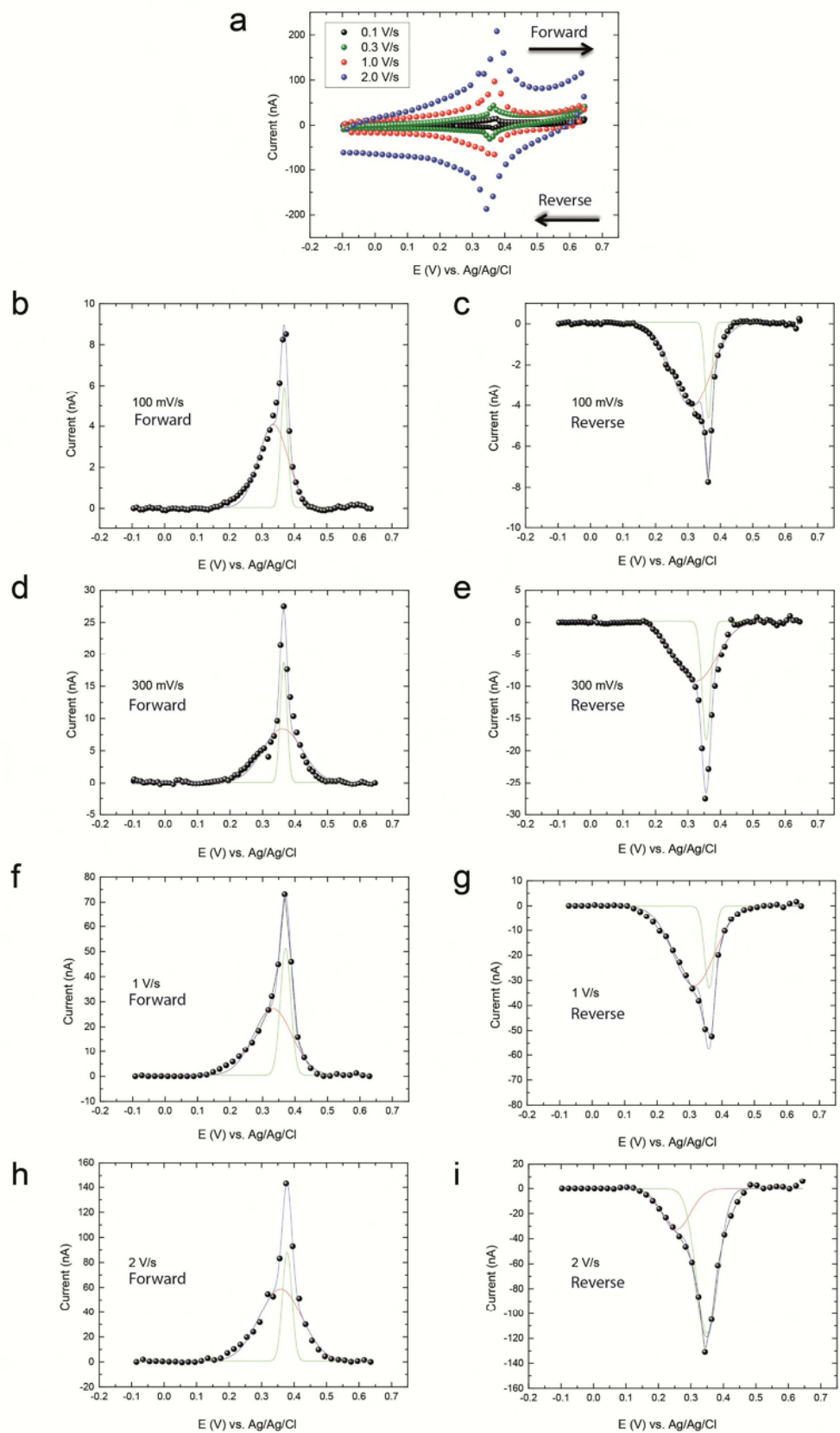

**Supplementary Figure 5 a** Raw datas for the cyclic voltammogram corresponding to Fig.1e, and voltammograms at different sweep rates (forward and reverse). **b-j** Cyclic voltammograms after removal of the base line (capacitive contribution) at different sweep rates (left:forward; right:reverse).

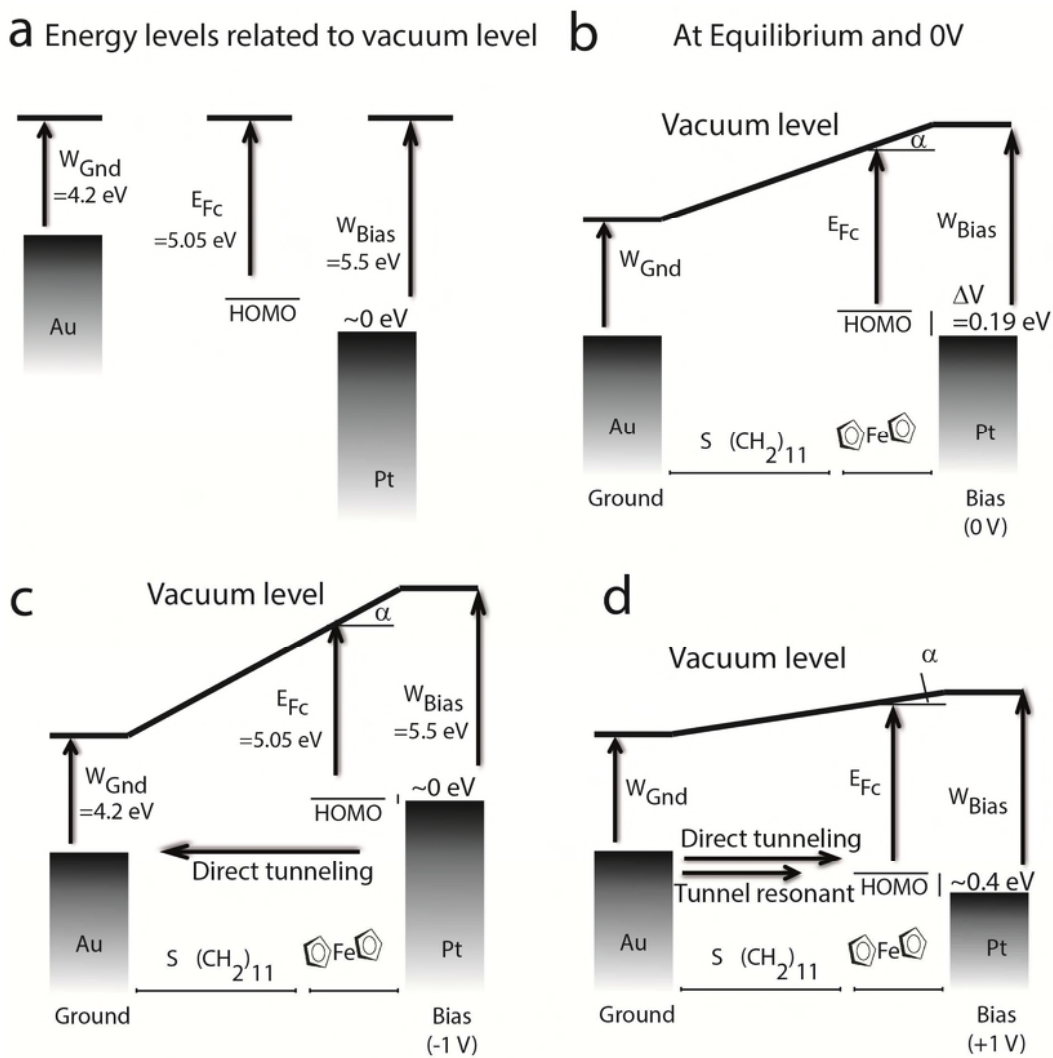

**Supplementary Figure 6 Construction of the energy band diagram for the molecular rectifier a** Energy levels related to vacuum for the electrodes (work functions) and for the molecules ($E_{HOMO}$, cf Supplementary Table 1 and Supplementary Notes 2). **b** When electrodes contact the molecule, at thermodynamic equilibrium and V=0V, Fermi levels of both electrodes are aligned. $\Delta E_{wf}$ represents the difference of metal work function. $\alpha\Delta E_{wf}$ represents the difference of energy between $E_{HOMO}$ and the gold electrode, that has to be taken into account due to the difference in electrodes metal work functions. Parameter $\alpha$=0.8 was considered as a typical value between $0.7^6$ and $0.9^7$. In ref.6, a GaIn electrode with a $Ga_2O_3$ oxide layer was used, and in 7, a Pt electrode was used. As parameter $\alpha$ can be defined as $V_r^2/(V_r^2+V_l^2)$ where $V_l$ and $V_r$ represent the energies of coupling to the left and right electrodes respectively, it is likely that the coupling between the right electrode and the Fc molecules is slightly reduced in the case of GaIn due to the presence of $Ga_2O_3$. By appling a bias on the Pt tip (-1V and +1V), the corresponding energy band diagrams are shown in **c** and **d**.

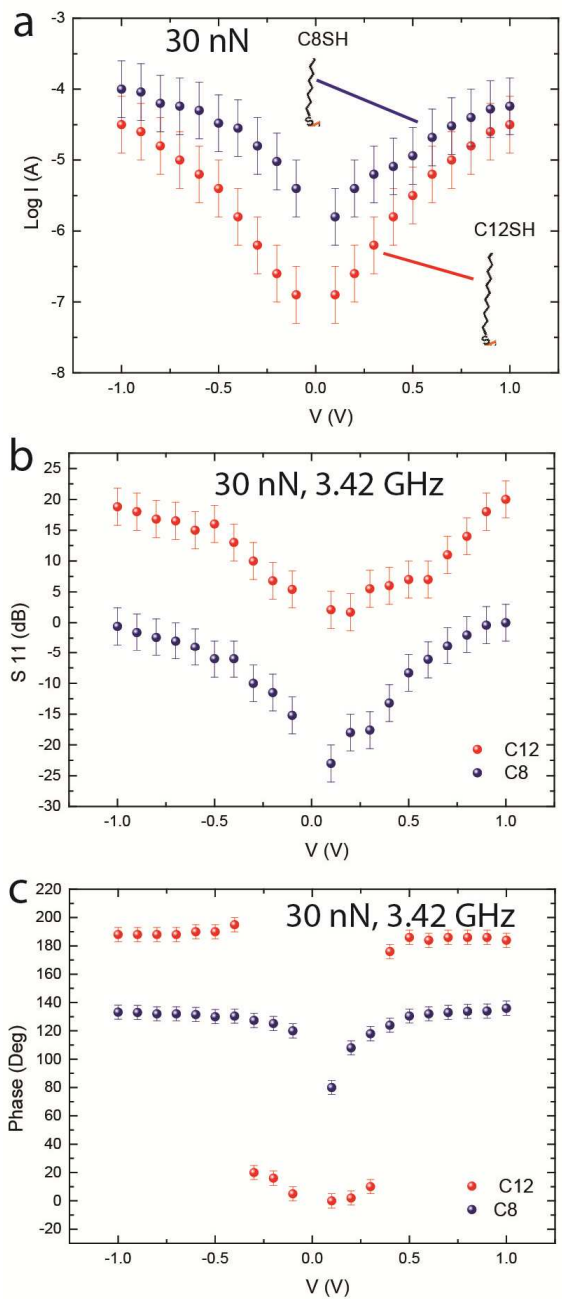

**Supplementary Figure 7 a** DC current (Log scale) for SAMs composed of $C_8$ and $C_{12}$ molecules. **b** Related $|S_{11}|$. **c** Related $S_{11}$ phase. In all curves, the dot corresponds to the maximum in the log normal distribution (from 100 dots) and the error bar is related to the standard deviation. The applied force was 30 nN. It is larger than the force applied on the $FcC_{11}$ because no variation of $S_{11}$-V could be observed at 18 nN for $C_{12}$ molecules (pure capacitive regime).

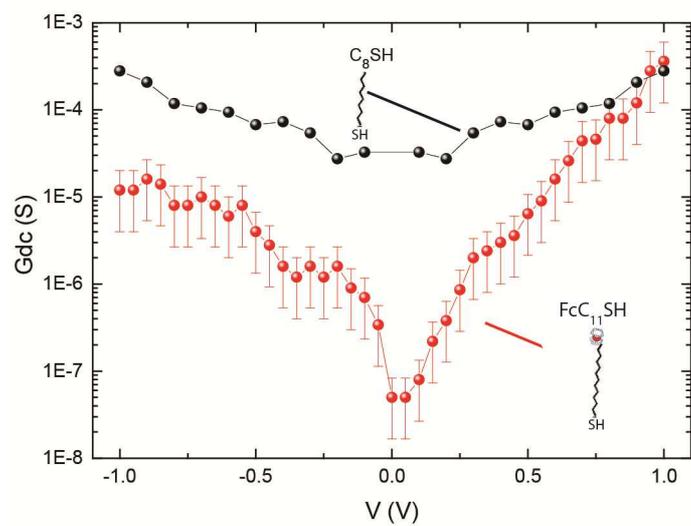

**Supplementary Figure 8** Dynamic conductance obtained from I-V curves for the $FcC_{11}SH$ (at 18 nN) and $C_8SH$ (at 30 nN) molecules

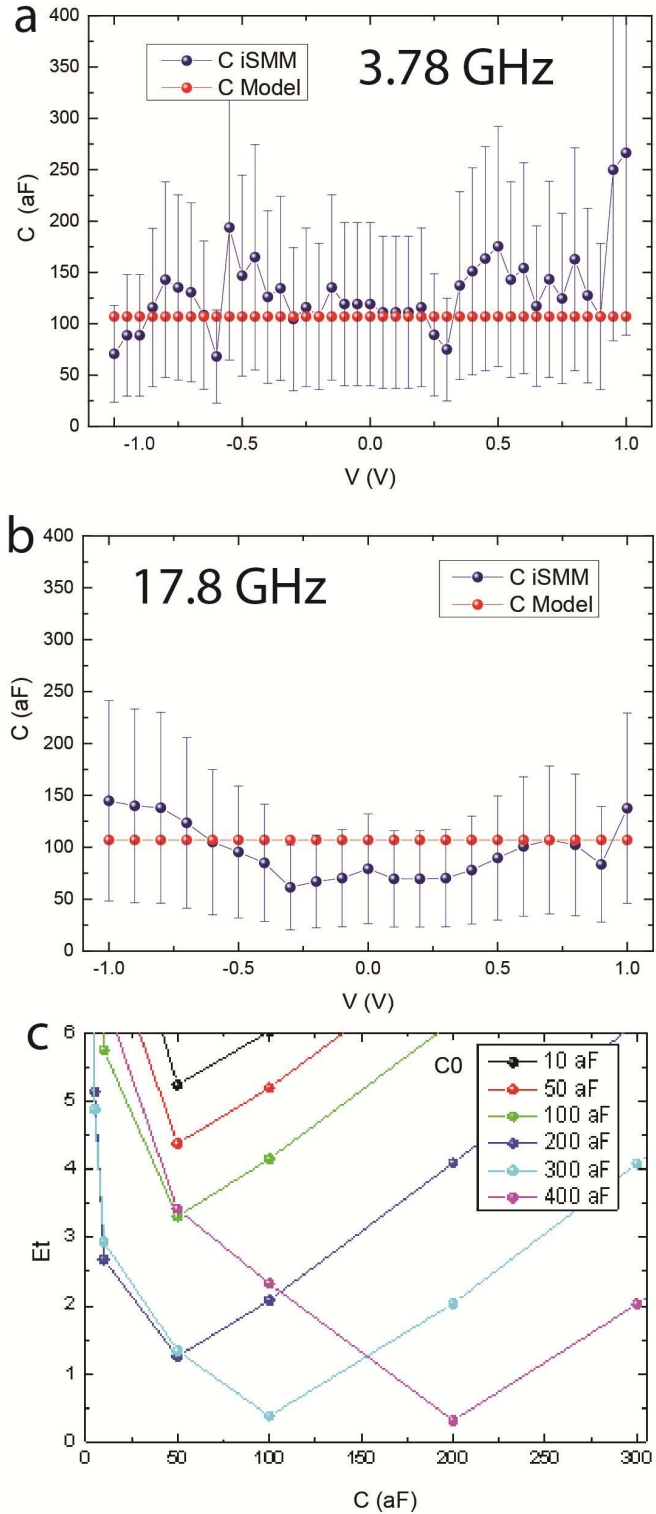

**Supplementary Figure 9 a** Estimated $C_{ISMM}$ from the 2 parameters model (61dB -153°) at 3.78 GHz. **b** Estimated $C_{ISMM}$ from the 2 parameters model (52.24dB -89.96°) at 17. 8 GHz. **c** Relative error in the fittings of conductance and capacitance using $S_{11}$ with respect the conductance measured by resiscope and the capacitance used in the model. Several values of $C_0$ are displayed.

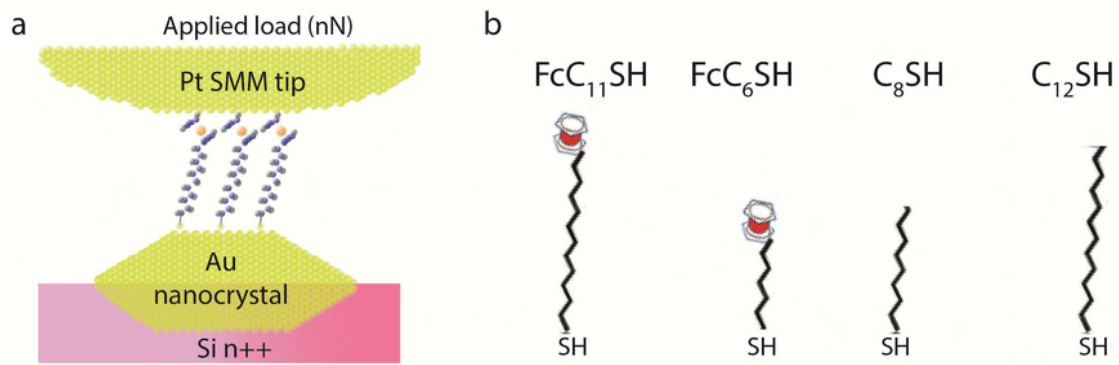

**Supplementary Figure 10 a** Schematic representation of the tip/molecules/Au nanocrystal structure showing that a load is applied on the tip. This parameter is tunable and its impact is discussed in that section. **b** Schematic representation of the four molecules tested in that study to evaluate the impact of molecule design on the performance of high frequency molecular electronics.

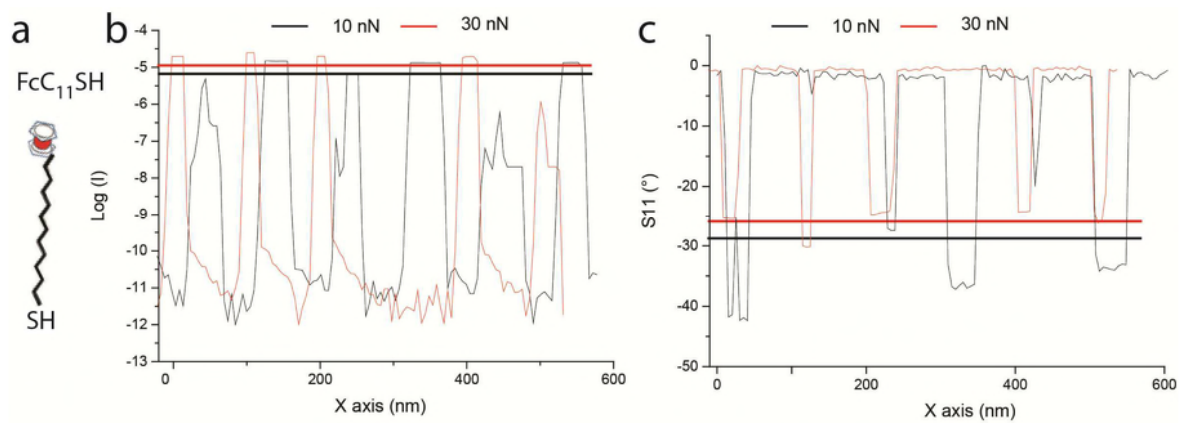

**Supplementary Figure 11 a** Schematic representation of the FcC$_{11}$SH molecule used in this study. **b** Raw datas of the measured current (log scale in A) on 6 dots at 0.8V for 2 different forces of 10 nN et 30 nN. **c** Raw datas of the measured S$_{11}$ Phase at 3.8 GHz, 0.8V on 6 dots for 2 different forces of 10 nN et 30 nN.

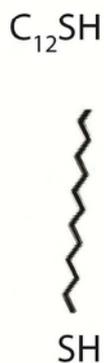
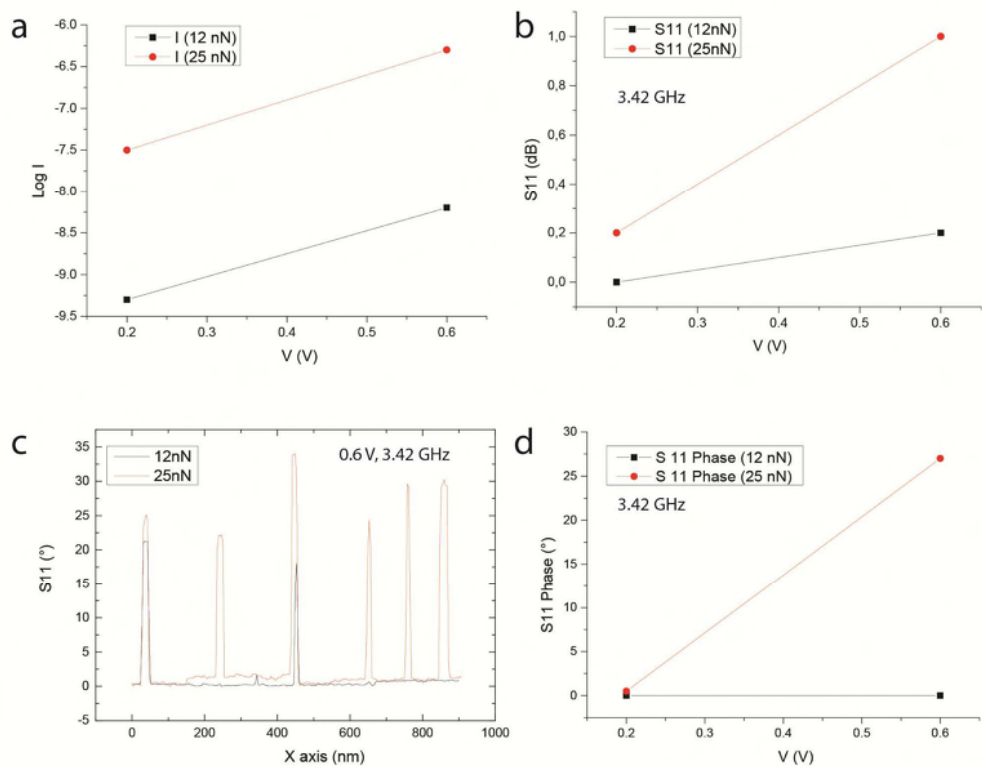

**Supplementary Figure 12** Left: Schematic representation of a $C_{12}SH$ molecule. **a** Evolution of the measured current for $C_{12}SH$ molecules on gold nanocrystals at two different biases and two different tip loads. **b** Evolution of the measured $|S_{11}|$ for $C_{12}SH$ molecules on gold nanocrystals at two different biases and two different tip loads (frequency is 3.42 GHz). **c** Raw datas of $S_{11}$ phase for 6 molecular junctions at two different loads (bias is 0.6 V and frequency is 3.42 GHz). The signal is very weak at 12 nN. **d** Evolution of the measured $|S_{11}|$ for $C_{12}SH$ molecules on gold nanocrystals at two different biases and two different tip loads (frequency is 3.42 GHz).

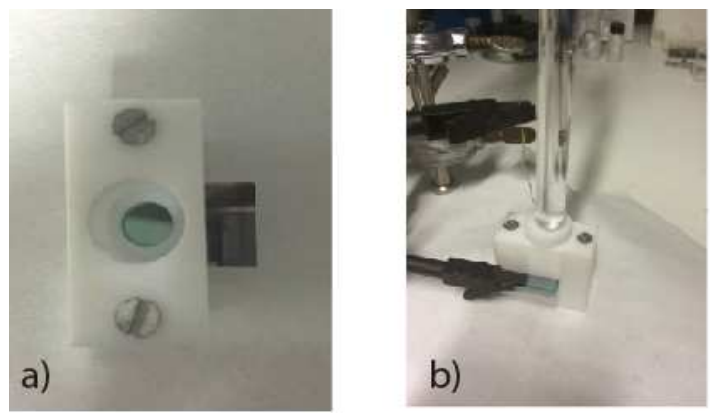

**Supplementary Figure 13. a** Electrochemical cell with a sample. **b** set up during the operation

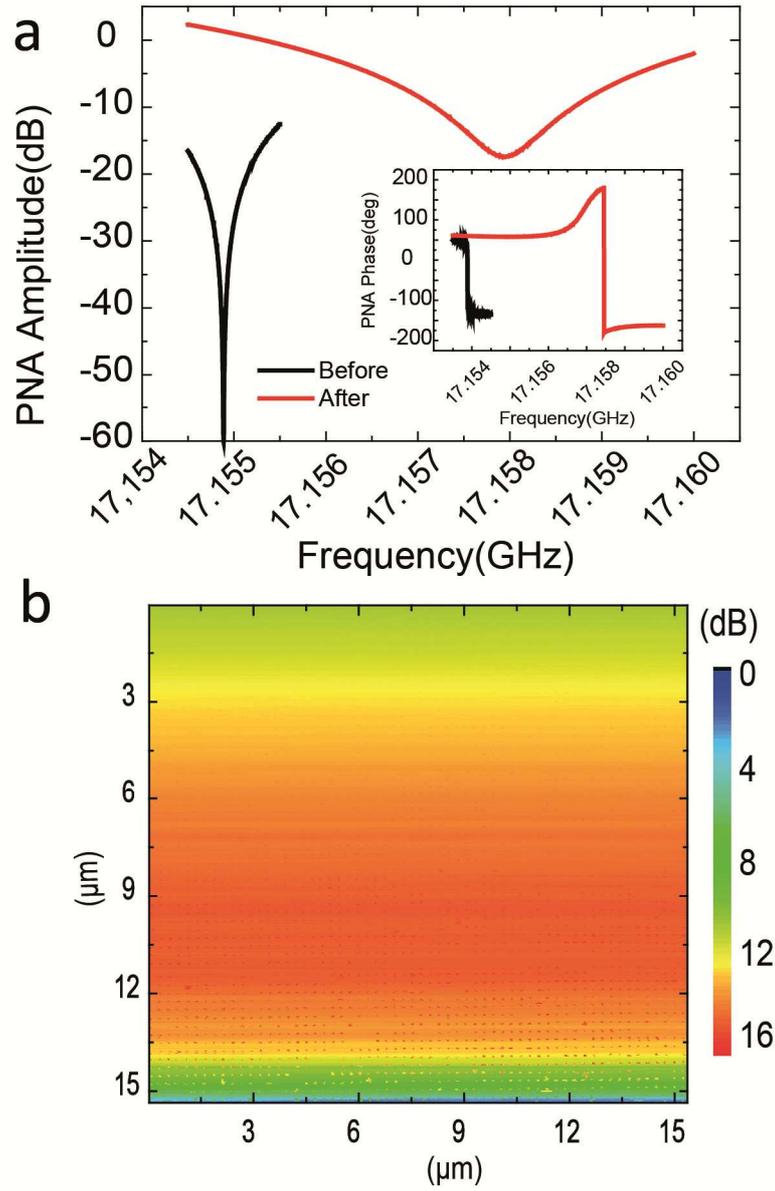

**Supplementary Figure 14. a** 17,157929 GHz spectrum before and after 1h 22 min. **b** $S_{11}$ amplitude image over 3600 nanojunctions.

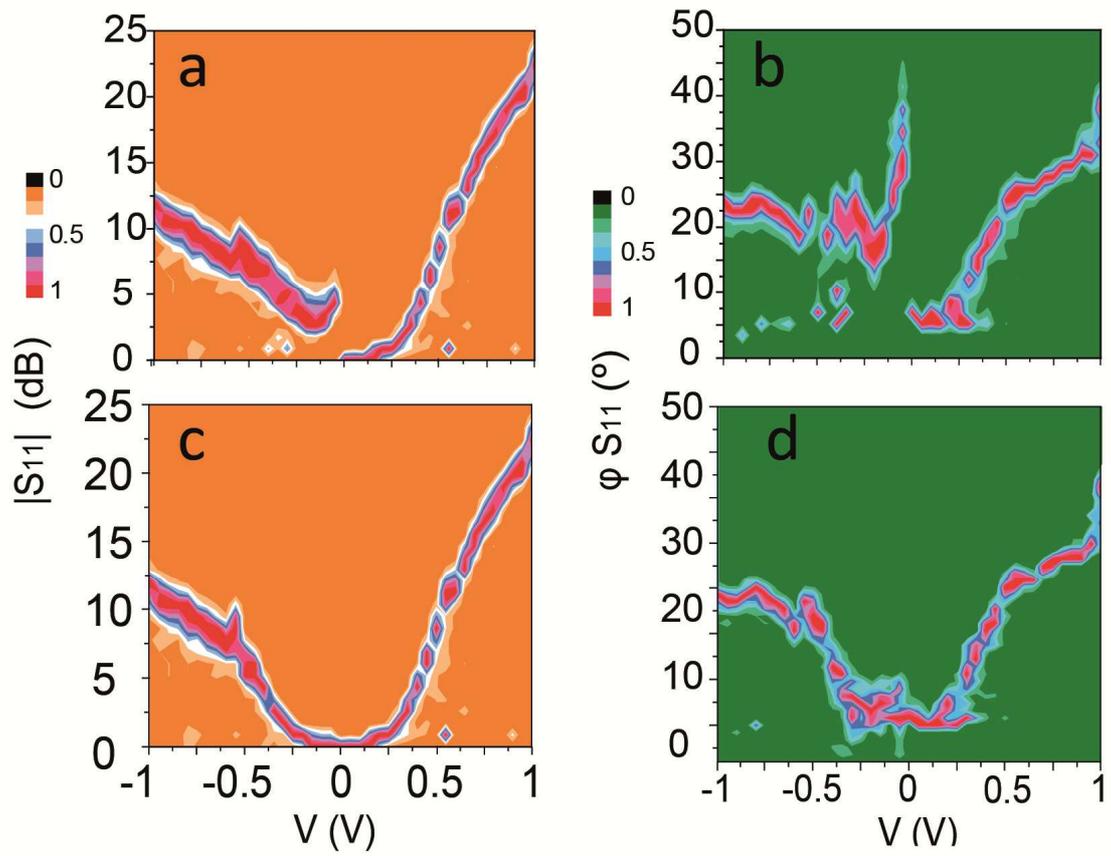

**Supplementary Figure 15. a** $S_{11}$ amplitude row data at 3.78 GHz, **b** phase row data, **c** $S_{11}$ amplitude corrected and **d** phase corrected.

**Supplementary Table 1:** Fitting parameters for cyclic voltammetry (100 mV/s)

| Model | | Gauss | |
|---|---|---|---|
| Equation | | y=y0 + (A/(w*sqrt(PI/2)))*exp(-2*((x-xc)/w)^2) | |
| Reduced Chi-Sqr | | 3,73E-20 | |
| Adj. R-Square | | 0,98846 | |
| | | Value | Standard Error |
| Peak1 | y0 | $2,33 \times 10^{-11}$ | $2,69 \times 10^{-11}$ |
| | xc | 0,3365 | 0,00158 |
| | w | 0,08718 | 0,00239 |
| | A | $4,44 \times 10^{-10}$ | $1,49 \times 10^{-11}$ |
| | sigma | 0,04359 | 0,00119 |
| | FWHM | 0,10264 | 0,00281 |
| | Height | $4,06 \times 10^{-9}$ | $1,20 \times 10^{-10}$ |
| Peak 2 | y0 | $2,33 \times 10^{-11}$ | $2,69 \times 10^{-11}$ |
| | xc | 0,36883 | $3,87 \times 10^{-04}$ |
| | w | 0,02085 | $9,40 \times 10^{-04}$ |
| | A | $1,52 \times 10^{-10}$ | $8,65 \times 10^{-12}$ |
| | sigma | 0,01042 | $4,70 \times 10^{-04}$ |
| | FWHM | 0,02455 | 0,00111 |
| | Height | $5,81 \times 10^{-09}$ | $2,08 \times 10^{-10}$ |

| Paper | Bottom Electrode /Wgnd (eV) | Top Electrode /Wbias (eV) | Junction area (cm$^2$) | Efc (eV) | α | J$_{max}$ @(-1V or 1V) (A/cm$^2$) | \|J(+1)/J(-1)\| | Rectification direction (sign of Vbias) | ΔV (eV) |
|---|---|---|---|---|---|---|---|---|---|
| Ref.16 | Ag/4.5 | EGaIn/4.3 | 3.14x10$^{-6}$ | 5.05 | 0.8 | 5.00x10$^{-4}$ | 151 | − | −0.71 |
| Ref.17 | Ag/4.3 | EGaIn/4.3 | 3.14x10$^{-6}$ | 5.05/4.8* | 0.8 | 1x10$^{-3}$/5x10$^{-3}$ | 610/1100 | − | −0.75/−0.5 |
| Ref.18 | Au/4.2 | PEDOT-PSS/5.05 | 2x10$^{-6}$ | 5.05 | 0.8 | 4x10$^{-7}$ | 1.5 | − | −0.17 |
|  | Au/5.1** |  |  |  |  |  |  | = | 0.01 |
| Ref.19 | Au/4.2 | Au NP/4.2 | ? | 5.05 | 0.5 | ? | 2*** | − | −0.85 |
|  | Au/4.8 | /4.8 |  |  |  |  |  |  | −0.25 |
| Ref.20 | Au/4.2 | STM tip/? | 3x10$^{-15}$**** | 5.05 | 0.8 | 1.6x10$^{+3}$ | 2 | + |  |
|  |  |  |  |  | 0.65 |  |  |  |  |
| Ref.21 | Au/4.2 | STM tip/5.1 | 3x10$^{-15}$**** |  |  | 1x10$^{+5}$ | 1.5 | − | −0.13 |
| This Work | Au/4.2 | Pt/5.5 | 3.14x10$^{-12}$ | 5.05 | 0.8 | 2x10$^{+7}$ | 5 | + | 0.19 |
|  | Au/4.8 |  |  |  |  |  |  | + | 0.31 |

\* Two types of configurations for Fc
\*\* Value considered in their paper
\*\*\* We considered their Fig.4c, e.g. a configuration as close as possible from other works
\*\*\*\* We considered only one molecule measured with the STM tip

**Supplementary Table 2** Comparison of properties of various molecular diodes using a Fc thiol monolayer

*Supplementary Note 1 Cyclic voltammogram analysis.*

Supplementary Figure 5a shows the raw data for the cyclic voltammogram measured at a scan rate of 0.1 V/s on an array of $5.1 \times 10^7$ nanodots (20 nm diameter) functionalized with $FcC_{11}SH$ molecules, and Supplementary Figure 5b shows the same data after correction from the capacitive contribution (the base line is removed using the baseline function from Origin software). The capacitive contribution is mainly due to the presence of a thin layer of silicon dioxide between dots. Lee et al.[8] reported an anodic peak deconvolution method using Gaussian and Lorentzian functions to probe the heterogeneity of the SAMs. In our CVs, we didn't notice a clear difference with the use of a Lorentzian function due to the sharp second peak, so peak deconvolution was performed with two Gaussian functions. Results from fits obtained with two Gaussian curves are shown in Supplementary Table 1. Supplementary Figures 5c-j show cyclic voltammograms (forward and backward) at four different sweep rates (100 mV, 300 mV, 1V and 2V). These 8 figures are used to extract an averaged density of molecules per dot of 0.405 +/- 0.04 $nm^2$/molecule (Supplementary Figure 4). The error bar is mainly related to the uncertainty on the area per dot.

For SAMs of $SC_nFc$ on Au, the theoretical value of 0.37$nm^2$/molecule (or a density of $4.5 \times 10^{-10}$ mol/$cm^2$) is usually considered assuming a hexagonal packing with Fc treated as spheres with a diameter of 6.7Å[9]. This value has been measured for the fully packed monolayers[10]. In this study, the averaged density of molecules on the gold nanocrystals is ~10% lower than for a fully packed monolayer, but in the range of values found for a dense monolayer. According to ref.10, the 10% lower density is observed at reduced van der Waals interactions, for example for a monolayer of $SC_8Fc$. In ref.10, they observed a new redox wave at low oxidation potentials for highly disordered SAMs when disulfides were present in the thiol precursors and a very broad redox waves when rough electrodes were used. Although we cannot exclude the presence of disulfides in the thiol precursor (we have not performed an additional purification step), their density may be small as such additional peak is not clearly observed in our cyclic voltammograms. Also, we did not observed broad voltammograms which may be explained by the small dispersion on gold nanocrystals structure.

From the CV curves, we can estimate the surface coverage $\Theta$, i.e. the number of molecules per dot and the area occupied per molecule using the following equation :

$$\Theta = Q_{tot}/nFA \qquad \text{Supplementary Equation 1}$$

where $Q_{tot}$ is the total charge involved during the redox process, n is the number of electrons transferred during the reaction (here 1), F the Faraday's constant (96 485 C/mol) and A the surface area. $Q_{tot}$ is given by the integral under the CV curve taken into account the scan sweep, v=0.1 V/s, we get $Q_{tot}=6\times10^{-9}$C from Fig. S5-b. The total area is the area of the nanodot multiplied by the number of dots. Considering a truncated cone of 20 nm diameter basis, 30 degree angle and 3 nm thick (Supplementary Figure 4c) gives an area of 72 nm$^2$ on the top part and 278 nm$^2$ on the sides, we obtain a surface of 350 nm$^2$ per dot. With $5.1\times10^7$ nanodots on a 500 µm x500µm (nanodot pitch 70 nm), A=$1.785\times10^{-4}$ cm$^2$. The estimated area per molecule of ~0.405 nm$^2$, corresponds to a number of FcC$_{11}$ molecules between the iSMM tip and the electrode of ~150 molecules.

*Supplementary Note 2 Estimation of $E_{HOMO}$ and Energy band diagram*

The energy of the HOMO, $E_{HOMO}$, relative to the vacuum is estimated from the voltammogram wave using:

$E_{HOMO} = -e(E_{1/2} + E_{Ag/AgCl-NHE}) + E_{abs,NHE}$     Supplementary Equation 2

with e the electron charge, $E_{abs,NHE}$ the absolute potential energy of the normal hydrogen electrode (NHE) at -4.5 eV[11], $E_{(Ag/AgCl-NHE)}$ is the potential of the Ag/AgCl versus the NHE (0.197 V). $E_{1/2}$ is the half-wave potential measured on the voltammograms. For a redox species immobilized on a surface, the anodic and cathodic potential are ~ equal (Supplementary Figures 5c,d), and $E_{1/2}$ is given by the peak position. Cyclic voltammograms are fitted considering two peaks at 0.37 V and 0.34V vs Ag/AgCl (Supplementary Table 1), so we consider only the level with larger $E_{HOMO}$. We obtain $E_{HOMO} \simeq -5.05$ eV as the related energy level in the molecular junction. Supplementary Figure 6 shows the proposed energy diagram of the molecular junction. We used theoretical values for the work functions of the Pt electrodes ($\approx 5.5$ eV[12,13]) and Au electrode (4.2 eV) following Ultraviolet Photoelectron Spectroscopy (UPS) results from ref.10 (Au with $FcC_{11}$ SAM). This schematic diagram is intended to explain the rectification observed for a positive bias applied on the Pt tip.

## Supplementary Note 3 Direction/strength of Rectification and current densities[14-19]

In Supplementary Table 2, we compare the difference of energy $\Delta V$ between the Ferrocene HOMO level and the energy level of the biased top electrode in a molecular junction configuration at 0V (Supplementary Figure 6) for seven different studies that used a FerrocenylAlkylthiol molecule, including the present work. The sign of $\Delta V$ provides a relevant indication on the direction of rectification (e.g. $\Delta V<0$ favors a diode "on" at negative bias, the usual case) and could be expressed analytically as:

$$\Delta V = (W_{bias} - E_{Fc}) - (1-\alpha)(W_{bias} - W_{gnd}) \quad \text{Supplementary Equation 3}$$

, where $W_{bias}$ and $W_{gnd}$ are the metal work functions of the biased and grounded electrodes respectively, $\alpha$ is the parameter of coupling between the Ferrocene groups and the biased electrode (in the Landauer formula[7], it is defined as $V_b^2/(V_g^2+V_b^2)$) where $V_b$ and $V_g$ represent the energies of coupling to the biased and grounded electrodes, respectively), and $E_{Fc}$ is the energy position of the Ferrocene HOMO level related to vacuum. For simple comparison and discussion, we first considered the same coupling parameter $\alpha=0.8$ for all studies (Supplementary Figure 6). Interestingly, the Supplementary Equation 3 leads to $\Delta V>0$ only in the case of the Pt biased electrode with large work function (Supplementary Table 2) used in this study. In the case of a top electrode with a relatively large work function such as PEDOT-PSS, $\Delta V\sim 0$ (a very small rectification at negative bias was observed). In other words, the difference in rectification direction in the present work compared to others could be mainly explained by the large work function of the Pt top electrode using an energy band diagram with a single energy level. The supplementary Equation 3 stresses the drastic importance of $W_{bias}$ if $\alpha$ gets close to 1. Nontheless, as clearly demonstrated in ref.20, the monolayer organization plays a key role in the "off" state, with large leakage (lower rectification) when the monolayers are not perfectly packed. We also expect that this plays a key role in addition to the metal work function.

Note that for RF molecular rectifier applications such as mixers, the figure of merit is the conductance in the "on" state as the non linearity close to this biasing point is exploited.

The current density has been extensively discussed in the main text. We can note that the current density for a single molecule junction (such as a STM tip top electrode) is still 2 orders of magnitude lower than that in the present study, which can be explained by the weak coupling between the STM tip and the Ferrocene molecules.

*Supplementary Note 4. Effect of the Tip load and the Molecule*

We have also investigated the effect of the tip load and the molecule types (Supplementary Figures 10-12). In the case of the nanocrystal structure, there is no increase of junction area with the tip load unlike for experiments performed on full substrate[21], but tip load are typically in the same range. For FcC$_{11}$SH molecules, we have noticed a weak effect of the tip load on the measured current and S$_{11}$ parameters (Supplementary Figure 11), unlike for C$_{12}$SH molecules (Supplementary Figure 12) where almost 2 orders of magnitude difference in current was observed when the load was changed from 12 nN to 25 nN. As mentioned on the main text, the weak effect of the load for FcC$_{11}$SH molecules could be a signature of weaker van der Waals interactions (reduced monolayer thickness). We have previously shown that short molecules such as C$_8$SH depend weakly on tip load, which is simply explained by the Hooke formula (small thickness change)[22], and deformation of electrodes.

In the case of strong van der Waals interactions or for thick monolayers, the effect of tip load provides another degree of freedom to tune f$_t$. We illustrate this for C$_{12}$SH molecules (Supplementary Figure 12). At low tip load (12 nN), we observe almost no effect of the tip bias on the S$_{11}$ phase as expected for a pure capacitor: $f_t$ is below 3.4 GHz on the whole voltage bias. Increasing the load to 25 nN enables to observe a difference in the phase and amplitude as expected for a tunnel junction with f$_t$>3.42 GHz at least for /V/>0.5 V. Considering a tip load of 30 nN to keep $f_t$ high enough for both C$_{12}$SH and C$_8$SH molecules, we confirmed that no clear rectification was observed neither in the current nor on /*S$_{11}$*/ (Supplementary Figure 7). Note that the current levels are in the same order of magnitude for the C$_8$SH and the FcC$_{11}$SH molecule in the "on" state, but the slope is larger for FcC$_{11}$SH molecule (Supplementary

Figure 8). Therefore, $FcC_{11}SH$ molecule is more appropriate for high performance RF diodes (e.g. mixer application: use of the nonlinearity at the biasing point in the "on" state).

We have also tried to study $FcC_6SH$, but we systematically reached unsustainable current densities for the gold nanocrystals.

## Supplementary Methods:

*S₁₁ model and calibration*

Under the hypothesis of a perfect interferometer, no parasitic reflection or mismatch among the components is taken into account. In the case of $Z_{DUT}$ (impedance of the device under test) much higher than $Z_c$ (characteristic impedance), the ratio $S_{11}$ between the wave transmitted after the interferometer to the receiver and the wave coming from the RF source can be approximately considered as proportional to the admittance of the device according to[23]:

$$S_{11} \approx A\Gamma + B \qquad \text{Supplementary Equation 4}$$

where $A$ and $B$ are two complex parameters. $\Gamma = \frac{Z-Z_C}{Z+Z_C}$ is the DUT reflection coefficient and $Z$ is the DUT impedance. For small DUT admittances $Y = \frac{1}{Z} \ll \frac{1}{Z_C}$, $\Gamma$ can be rewritten as $\Gamma \approx 1 - 2Z_C Y$ and the Supplementary Equation 4 becomes:

$$S_{11} \approx -2Z_C A(Y - Y_0) \qquad \text{Supplementary Equation 5}$$

where $Y_0 = \frac{A+B}{2Z_C A}$. The complex factor "$A$" takes into account losses, gain and shifts due to cables, passive and active elements. $Y_0$ is the admittance for which fully destructive interference occurs ($S_{11}=0$).

In order to fit the measured $S_{11}$, we considered an equivalent electrical circuit composed of a resistance and capacitances (intrinsic and extrinsic) in parallel.

The measured admittance Y will therefore be modeled by equation:

$$Y = G + C\omega j \quad \text{Supplementary Equation 6}$$

where ω is the angular frequency ($2\pi f$).

$G$ is the (dynamic) conductance that can be derived in a first approach from the derivative of the (DC) *I-V* curve due to the absence of photocurrent, C is given by $C_{mol}+C_p$.

In order to use the Supplementary Equation 5, we need to identify *A* and $Y_0$. To do that, we need to perform two measurements of $S_{11}$ and use the corresponding values of *G* and *C* as a model. First, the interference is set for $S_{11}=0$ and is measured when the tip is on the SiO$_2$ surface (between the dots). Then the two points are chosen for given DC voltages among the measurements on the molecules to obtain the best fit. *G* will be obtained from the DC conductance and we assume that *C* does not change with the voltage. *C* is estimated as the value that gives the best fitting when comparing the conductance measured at DC and the conductance extracted from $S_{11}$. The error $E_t$ is calculated as the sum of the fitting error in capacitance and conductance obtained by equations (Supplementary Figure 9) where $C_M$ and $G_R$ are respectively the modeled capacitance and the conductance measured by the resiscope and $C_{iSMM}$ and $G_{iSMM}$ are the capacitance and conductance extracted from S$_{11}$:

$$E_C = \frac{C_M - C_{iSMM}}{C_M} \quad \text{Supplementary Equation 7}$$

$$E_G = \frac{G_R - G_{iSMM}}{G_M} \quad \text{Supplementary Equation 8}$$

$$E_t = E_C + E_G \quad \text{Supplementary Equation 9}$$

We see that when we go from, for example, C$_0$= 350 aF to C$_0$= 470 aF the minimum in the relative error of the conductance move from 42% to 20%. We can see how all our variables have a relationship. $C_0$=300aF is close to the calibrated value measured on the same wafers and same SMM tips and it is a reasonable value looking at the literature[24]. In Supplementary Figure 9 we can see that the corresponding values of nano molecular junctions that minimizes the error is around 107 aF. This value is still far from the supposed value of few aF corresponding to the molecules alone.

# Supplementary References: